\documentclass[english,prd,onecolumn,nofootinbib]{revtex4}
\pdfoutput=1
\usepackage{graphicx, bm, color, babel}
\usepackage{hyperref}
\usepackage{amsmath}
\usepackage{wrapfig}
\usepackage{color}

\def\be{\begin{equation}}
\def\ee{\end{equation}}
\def\bea{\begin{eqnarray}}
\def\eea{\end{eqnarray}}

\def\d {\mathrm{d}}
\newcommand{\HH}{\mathcal{H}}
\newcommand{\HA}{\widehat{\mathcal{H}}}
\newcommand{\HB}{\overline{\mathcal{H}}}

\renewcommand{\>}{\rangle}
\newcommand{\sfrac}[2]{{\textstyle\frac{#1}{#2}}}

\renewcommand{\o}{{\hspace{-0.5mm}{\text{\tiny$\perp$}}}}
\newcommand{\p}{_{{\text{\tiny$\|$}}}}

\newcommand{\J}{{\bm{\mathcal{J}}}}
\newcommand{\K}{{\bm{\mathcal{K}}}}
\newcommand{\I}{{\bm I}}

\newcommand{\F}{{\bm F}}
\renewcommand{\S}{{\bm S}}
\newcommand{\x}{{\bm \xi}}
\renewcommand{\a}{{\bm \alpha}}
\renewcommand{\b}{{\bm \beta}}
\newcommand{\R}{{\bm{\mathcal{R}}}}
\newcommand{\A}{{\bm{\mathcal{A}}}}
\renewcommand{\P}{{\bm{\mathcal{P}}}}

\begin{document}

\title{The general theory of secondary weak gravitational lensing }
\author{Chris Clarkson\\
\emph{Astrophysics, Cosmology \& Gravity Centre, and, Department of Mathematics \& Applied Mathematics, University of Cape Town, Cape Town 7701, South Africa.}}

\date{\today}

\begin{abstract}

Weak gravitational lensing is normally assumed to have only two principle effects: a magnification of a source and a distortion of the sources shape in the form of a shear. However, further distortions are actually present owing to changes in the gravitational field across the scale of the ray bundle of light propagating to us, resulting in the familiar arcs in lensed images. This is normally called the flexion, and is approximated by Taylor expanding the shear and magnification across the image plane. However, the physical origin of this effect arises from higher-order corrections in the geodesic deviation equation governing the gravitational force between neighbouring geodesics~-- so involves derivatives of the Riemann tensor. We show that integrating the second-order geodesic deviation equation results in a `Hessian map' for gravitational lensing, which is a higher-order addition to the Jacobi map.  We derive the general form of the Hessian map in an arbitrary spacetime paying particular attention to the separate effects of local Ricci versus non-local Weyl curvature. We then specialise to the case of a perturbed FLRW model, and give the general form of the Hessian for the first time. This has a host of new contributions which could in principle be used as tests for modified gravity.

\end{abstract}

\maketitle

\section{introduction and overview}

Weak gravitational lensing is becoming an important cosmological probe. The usual weak gravitational lensing theory depicts that a lens induces a convergence (a spin 0 mode) and a shear (a spin 2 mode) to a source lying behind it. There are two ways to describe this for a given mass distribution: one is to calculate all the null geodesics converging at an observer, $k^a$, and examine the output. Another is to calculate the propagation of a geodesic deviation vector, $\xi^a$, and examine the invariant moments of the resulting image distortion. This results in the Jacobi map between a source and an image~\cite{Seitz:1994xf,Perlick:2004tq}. The `weak lensing' calculation route typically takes the second option, as it produces accurate results easily and intuitively. The computation of the convergence and shear is achieved from the geodesic deviation equation which is linear in the deviation vector:\footnote{$a,b,c,\cdots$ denote spacetime indices, $A,B,C,\cdots$ are tetrad indices in the screen space,  $k$ or $\xi$ as an index denotes projection of that index in the direction of $k$ or $\xi$. $R^a_{~bcd}$ is the Riemann tensor. A dot is a derivative along the null curve~-- full details below.}
\be\label{sdjbcsbds}
\ddot \xi^a + R^a_{~k b k}\xi^b=0\,.
\ee
 Yet progressively stronger lensing events produce arcs and other more complicated distortions which cannot be captured by a simple convergence plus shear distortion. How can these be described within the weak lensing formalism?

The geodesic deviation equation is of course linear by construction: the rhs of \eqref{sdjbcsbds} should read $\mathcal{O}(\xi,\dot\xi)^2$, for the terms that are ignored. More complicated lensing events can therefore be described by examining this equation to higher order. Up to second-order in $\xi,\dot\xi$ we have the Bazanski equation~\cite{1977AnIHP..27..145B,1977AnIHP..27..115B,Vines:2014oba,schutz}:
\be\label{lasdbasvdb}
\ddot \xi^a + R^a_{~k b k}\xi^b + \nabla_{(k}R^a_{~b)ck} \xi^b\xi^c + 2R^a_{~bck} \xi^b\dot\xi^c = \mathcal{O}(\xi,\dot\xi)^3\,.
\ee
Naturally, derivatives of Riemann induce higher-order changes in the deviation vector. In this paper we extend the general weak lensing formalism to include all second-order corrections in the deviation vector. We aim to give the general solution to the Bazanski equation in an arbitrary spacetime, in terms of a `Hessian map' $\mathcal{H}_{ABC}$, which is the higher-order equivalent of the Jacobi map~$\mathcal{J}_{AB}$:
\be
\xi_A=\mathcal{J}_{AB}\zeta^B+\mathcal{H}_{ABC}\zeta^B\zeta^C\,
\ee
where $\zeta^A$ is an angle between neighbouring rays at the observer, and $\xi^A$ is the same at the source. 
 The Jacobi map is a rank 2 tensor in the image plane and, as such, has two spin 0 (a convergence from the trace and a rotation from the anti-symmetric part) and spin 2 degree of freedom (the shear from the trace-free symmetric part).  The Hessian map is a rank 3 tensor in the image plane, has a symmetry on two indices, so can be irreducibly decomposed into two vectors and a symmetric trace-free rank 3 tensor. In general, these correspond to two spin 1 modes, and a spin 3 mode (each having two polarisations). 

Secondary lensing effects have been considered in the past, and generically go under the name `flexion', describing classic arc shaped images, though there are other effects from secondary lensing~\cite{Goldberg:2004hh,Bacon:2005qr,Castro:2005bg,Okura:2006fi,Goldberg:2006jp,Massey:2006du,Schneider:2007ks,Bernstein:2008ah,Bacon:2008zj,Lasky:2009ca,Bacon:2009aj,Leonard:2009fa,Hilbert:2010am,Er:2010nt,Pires:2010ar,Munshi:2010cz,Fluke:2011ej,Schaefer:2011ui,Camera:2011me,Rowe:2012ih}. In these works, the flexion has been derived by Taylor expanding the amplification matrix (which is usually the linearisation of the Jacobi map about an FLRW background) across the image plane. In the notation above, this implies
\be\label{sdjkckjncsc}
\mathcal{H}_{ABC}\sim\frac{1}{2}\nabla_{(C}\mathcal{J}_{|A|B)}\,.
\ee 
This gives an approximation to the full Hessian, involving screen derivatives of the convergence and shear. These are in fact the leading contributions in most situations, because they capture the changing gravitational field across the image place, integrated along the line of sight. 

The leading contribution to $\mathcal{H}_{ABC}$ in a perturbed Minkowski background can be derived easily. The leading term is the one with the largest number of screen-space derivatives, and comes from the third term in~\eqref{lasdbasvdb}
\be
\nabla_{(k}R^a_{~b)ck} \xi^b\xi^c
\sim \frac{1}{2}\xi^b\xi^c\nabla_{b}R^a_{~kck} 
\sim \frac{1}{2}\xi^b\xi^c \partial_b\partial_c\Gamma^a_{~kk}
\sim -\frac{1}{4}\xi^b\xi^c\partial_b \partial_c \partial^a \delta g_{kk}
\sim  \xi^b\xi^c \partial_b\partial_c \partial^a \Phi\,,
\ee
where all terms without 3 screen space derivatives have been ignored. Integrating and projecting everything into the screen space then gives
\be
\mathcal{H}_{ABC}\sim 
\chi^2
\nabla_{A}\nabla_{B}\nabla_C
\int_0^\chi\d\chi'\frac{\chi(\chi'-\chi)}{{\chi'}}\Phi\,,
\ee
which is the 3rd derivative of the usual lensing potential. This is the same as can be derived from~\eqref{sdjkckjncsc}. However, as we shall see there are many more contributions to the Hessian than given by this approximation.

In this paper we make the link between flexion and geodesic deviation for the first time. In doing so we derive the general form of the Hessian, valid in any spacetime, which contains many more subtle contributions beyond the approximation~\eqref{sdjkckjncsc}. We shall then specialise to the case of perturbed FLRW model.

\section{Description of a null curve and the screen space}

\begin{wrapfigure}[25]{r}[0pt]{0.4\textwidth}
\begin{center}
\includegraphics[width=0.4\textwidth]{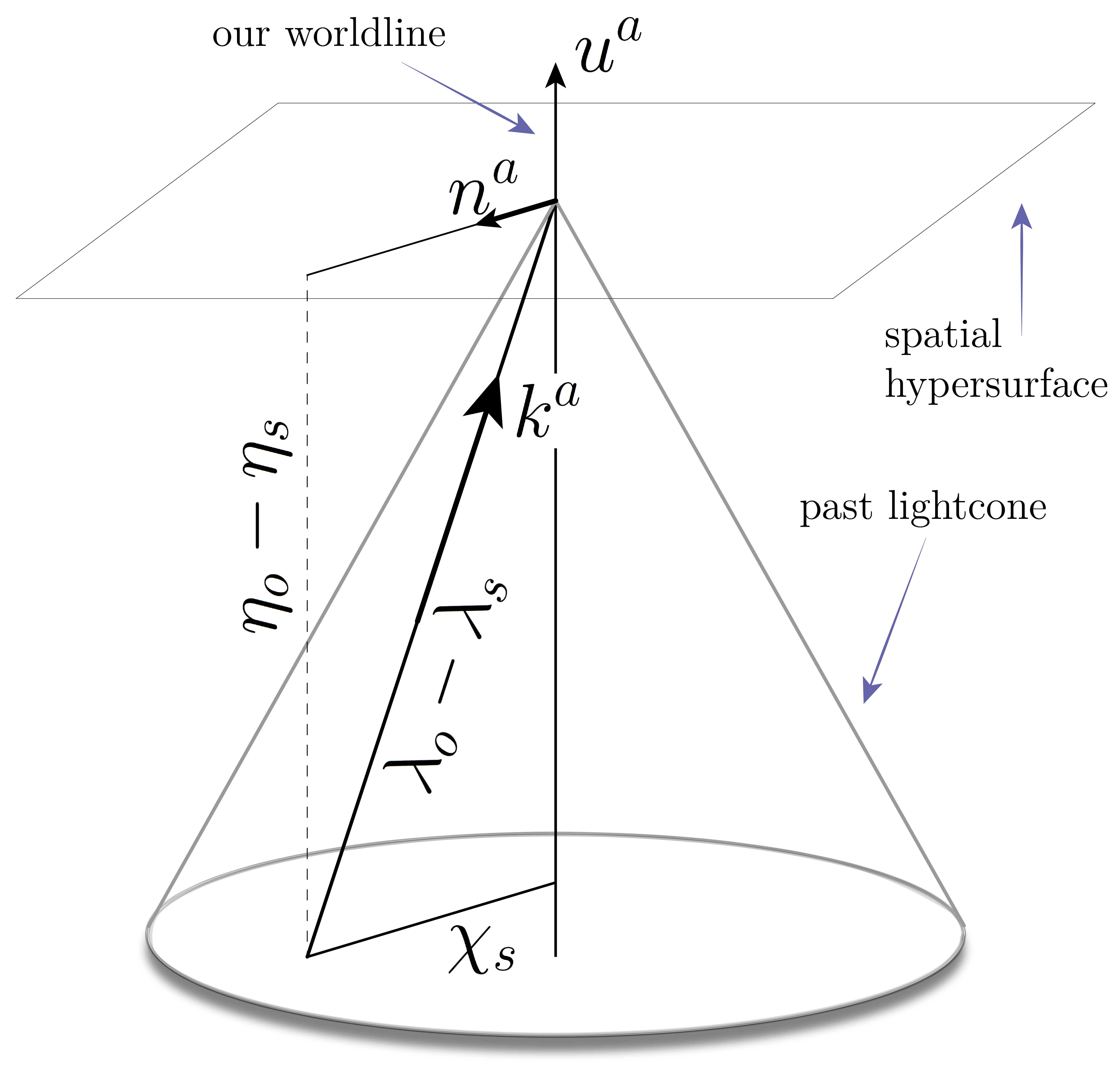}
\caption{Spacetime diagram showing our definition of 4-velocity and null vector. $\chi$ and $\eta$ are used in Sec.~\ref{kjdsskjdvbkdjvbs}.}
\label{spacetime}
\end{center}
\end{wrapfigure}

Consider a light ray with tangent vector $k^a$ and affine parameter~$\lambda$ on the past light cone which is  a  constant phase hypersurface, ${S}=\,$const :
 \be
{ k}^{a} = \frac{\d { x}^{a}}{\d\lambda} \,,~~~{k}_a= {\nabla}_a  S.
 \ee
The tangent vector is null and geodesic:
\begin{equation}\label{Geodesiceqn1}
{k}_a{k}^a=0,\,\,\,\,\,  {k}^b{\nabla}_b{ k}^a=0\,,
\end{equation}
and may be decomposed relative to an observer with  4-velocity $u^a$ into parallel and orthogonal components:  
\begin{eqnarray}\label{4vector}
{ k}^a=(-{ u}_b{ k}^b)\left({ u}^a-{ n}^a\right)={E}\left({ u}^a-{ n}^a\right),\nonumber\\~~~n_an^a=1,~n_au^a=0, ~{E}=-{ u}_b{ k}^b.
\end{eqnarray}
Here $n^a$ is the unit direction vector of observation, and $k^a$ is along an \emph{incoming} light ray on the past light cone of the observer. $E$ is the photon energy measured by $u^a$.  Note that our choice of $n^a$ is {\em opposite} to the direction of photon propagation.  

The screen space is orthogonal to the light ray and to the observer 4-velocity, and the tensor 
\begin{equation}\label{Screenspacemtric}
{N}_{ab}= {g}_{ab}+{u}_a{u}_b-{n}_a{n}_b\,,
\end{equation}
projects into screen space. It
satisfies the following relations
\begin{equation}
 {N}^a{}_a=2,
~~~ {N}_{ac}{{N}^c}_b={N}_{ab},
~~~{N}_{ab}{k}^b={N}_{ab}{u}^b={N}_{ab}{n}^b=0\,.
\end{equation}

Given the 4-velocity $u^a$ we can invariantly decompose tensors into scalars, vectors and projected, symmetric and trace-free tensors~\cite{Ellis:1998ct,Maartens:1996ch}, which are all spatial. Then,
For any  spatial tensor $T^{a\cdots }{}_{\cdots b}$, we can isolate the parts lying in the screen space and parallel to $n^a$ in a 1+1+2 decomposition~\cite{Clarkson:2002jz,Clarkson:2007yp}: 
\begin{eqnarray}
{{T}_\o}^{a\cdots }{}_{\cdots b}&=&{N}^a{}_c\cdots {N}^d{}_b {T}^{c\cdots }{}_{\cdots d}\,,\\
{T}\p&=&{n}_a\cdots n^b\,{T}^{a\cdots }{}_{\cdots b}\,.
\end{eqnarray}
For a PSTF tensor, we use $\|$ to represent all the indices which are projected along $n^a$. Traces in the screen space can also be removed. For example, a rank 2 PSTF tensor has the full decomposition 
\be
X_{ab}=X_{\langle
ab\rangle}=X\p\left({n_an_b-\frac{1}{2}N_{ab}}\right)+2{X\p}_{(a}n_{b)}+\left({N_{(a}^{~~c}N_{b)}^{~~d}-\frac{1}{2}N_{ab}N^{cd}}\right)X_{cd}\,.
\label{tensor-decomp} 
\ee

The invariant decomposition of the covariant derivative of the photon ray vector is given by,
\begin{equation}
 {\nabla}_b k_a= \frac{1}{2}\theta N_{ab}+ {\sigma}_{ab}\,,
\end{equation}
where 
\begin{eqnarray}
{\theta}= N^{ab} {\nabla}_a k_b,~~~~{\sigma}_{ab}= N_{(a}{}^c N_{b)}{}^d {\nabla}_c  k_d-\frac{1}{2}{\theta}N_{ab}\,.
\end{eqnarray}
Thus  ${\theta}$ describes the rate of expansion of the area of a bundle of light rays and ${\sigma}_{ab}$ describes its rate of shear (the trace-free part of the derivative projected into the screen space).  Note that there is no null vorticity since $k_a={\nabla}_aS$.

The covariant derivative of the 4-velocity is invariantly decomposed as:
 \begin{eqnarray}
 {\nabla}_{b} u_{a}=- A_{a}u_{b}+\frac{1}{3} \Theta ( g_{ab }+u_au_b) 
+ \Sigma_{ab}+  \Omega_{ab}\,,
 \label{du}
\end{eqnarray}
where $ \Theta$ is  the volume expansion rate of the $u^a$ worldlines,   $ A_a$ is the 4-acceleration,   $ \Sigma_{ab}$ is the shear tensor and $\Omega_{ab}$ is the vorticity tensor. 
In terms of these variables, the 1+1+2 decomposition of the covariant derivatives of $E$ and $n^a$ are:
\bea
E^{-1}\nabla_aE&=& -A\p u_a+\frac{1}{3}\Theta n_a+{\Sigma\p}_a+{\Omega\p}_a\\
\nabla_b n_a &=&-A\p (u_au_b- u_bn_a) +\frac{1}{3}\Theta (u_a n_b-n_an_b)+
(u_a - n_a) ({\Sigma\p}_b+{\Omega\p}_b) - u_b A_a \nonumber\\&&
+\left(\frac{1}{3}\Theta-\frac{1}{2}\frac{\theta}{E}\right)N_{ab}
-E^{-1}\sigma_{ab}+ \Sigma_{ab}+  \Omega_{ab}
\eea

The Sachs propagation equations for the null shear and null expansion are found from the Ricci identities~\cite{Sachs:1961zz,2013GReGr..45.2691J,1966RSPSA.294..195B,Clarkson:2011br},
\begin{eqnarray}\label{Raychau}
\frac{\d{\theta}}{\d \lambda}&=&-\frac{1}{2}{\theta}^2-  {\sigma}_{ab}\sigma^{ab}- R_{ab} k^a k^b\,, \\
\label{eq:shearevo}
\frac{\text{D}{\sigma}_{ab}}{\d\lambda}&=&-{\theta}{\sigma}_{ab}+ C_{acbd}  k^c k^d\,,
\end{eqnarray}
where ${\text{D} }/{\d\lambda}= k^a{ \nabla}_a $.
The photon geodesic equation~(\ref{Geodesiceqn1}) reduces to the equations for the photon  energy and  observational direction  
\begin{eqnarray}\label{energyprorpa1}
\frac{\d  E}{\d \lambda}&=&- E^2\left[\frac{1}{3}\Theta -  A_{a}n^{a}+ \Sigma_{ab} n^{a} n^{b}\right]\,,
\\
\frac{\text{D}n^a }{\d\lambda}&=&E \left[n^a\left(A_bn^b-\Sigma_{bc}n^bn^c \right) -A^a+ \left(\Sigma^a{}_b-\Omega^a{}_b\right) n^b
+ u^a\left(\frac{1}{3}\Theta - A_bn^b + \Sigma_{bc}n^bn^c\right)\right].
\end{eqnarray}

We shall project vectors and tensors onto the screen space using the tetrad basis $e_A^{~~b}$, $A,B=1,2$ where 
\be
e_A^{~~b}u_b=e_A^{~~b}n_b=0,~~~N_{ab} e_A^{~~a}e_B^{~~b}=g_{ab} e_A^{~~a}e_B^{~~b}=\delta_{AB},~~~
N_{ab} \dot e_A^{~~b}=0.
\ee
The derivatives of the tetrad (Ricci rotation coefficients) are not required here.

\section{the Null Geodesic Deviation equation to second-order}

Consider a deviation vector $\xi^a$ lying in the screen space orthogonal to $k^a$. This links two neighbouring geodesics, one at $x(\lambda)$ and the other at $x'(\lambda)$. The deviation vector at $x(\lambda)$ is defined as the covariant derivative of Synge's world function, which is half the squared proper length of the unique geodesic connecting $x$ and $x'$~-- see~\cite{Vines:2014oba}.  To second-order this obeys a generalisation of the GDE, known as the Bazanski equation ($\dot~=D/d\lambda$):
\be\label{kjsdbncsn}
\ddot\xi^a+k^b\xi^ck^d{R^a}_{bcd}
+\frac{1}{2}\xi^b k^c \xi^d k^e \nabla_eR_{bc~d}^{~~a}
+\frac{1}{2}\xi^b k^c k^d \xi^e \nabla_eR_{bc~d}^{~~a}
+2\dot\xi^b\xi^ck^d{R^a}_{bcd}=\mathcal{O}(\xi,\dot\xi)^3
\ee
An additional term $\frac{2}{3} \dot\xi^b\xi^c\dot\xi^d{R^a}_{bcd}$ would make this accurate in all powers of $\dot\xi$, but we shall not include this term here. Our initial aim is to give the solution to this equation.

We shall project this onto the screen space using the tetrad basis $e_A^{~~b}$. The tetrad components of $\xi^a$ in the screen space are $\xi_A=e_A^{~b}\xi_b$. This is called the Sachs basis.  The full connecting vector is
\be
\xi_a=e_a^{~B}\xi_B+\xi_k k_a\,.
\ee
This has a part parallel to $k^a$ at second-order which cannot be set to zero as it can at first. This part obeys
\be
\ddot\xi_k=-\frac{1}{2}\xi^A\xi^B \dot R_{AcBd}k^ck^d
+2\dot\xi^A\xi^B R_{AcBd}k^ck^d\,.
\ee
Thus, the deviation vector is forced out of the screen space as it is transported along $k^a$. However, since $\xi_k$ is second-order, it can only influence the screen parts of $\xi^a$, $\xi^A$, via the linear term in the GDE~-- but by the symmetries of the Riemann tensor, this contribution is zero. We therefore do not consider $\xi_k$ any further.

Then, the screen-projected part of \eqref{kjsdbncsn} can be written as 
\be\label{dsjhbcsbc}
\ddot\xi_A=\mathcal{R}_{AB}\xi^B+\mathcal{P}_{ABC}\xi^B\xi^C+\mathcal{Q}_{ABC}\xi^B\dot\xi^C\,,
\ee
where
\bea
\mathcal{R}_{AB}&=&-R_{ABc}k^c=\mathcal{R}_{BA}\,,\\
\mathcal{P}_{ABC}&=&-\frac{1}{2}\dot R_{BAC}
-\frac{1}{2}e_A^{~~a}e_{(B}^{~~b}e_{C)}^{~~c}\nabla_c(R_{adbe}k^dk^e)\,,\\
\mathcal{Q}_{ABC}&=&\frac{1}{2}R_{ABC}-\frac{3}{2}R_{BAC}\,.
\eea
Here we have defined the Riemann tensor with one index projected onto $k^a$ as
\bea
R^a_{~cd}&=&R^{ab}_{~~~cd}k_b\\
&=& C^{ab}_{~~~cd}k_b +\delta^a_{~[c}R^b_{~d]}k_b - k_{[c}R^a_{~d]}
-\sfrac{1}{3}R \delta^a_{~[c}k_{d]}\,,
\eea
where $C^{ab}_{~~cd}$, $R_{ab}$ and $R$ are the Weyl tensor, the Ricci tensor, and Ricci scalar. Then,
\bea
\mathcal{R}_{AB}&=& -C_{AcBd}k^ck^d -\frac{1}{2}\delta_{AB} R_{cd}k^ck^d\,,\\
{R}_{ABC}&=&C_{AdBC}k^d +\delta_{A[B}R^b_{~C]}k_b\,.
\eea
$\mathcal{R}_{AB}$ is often called the optical tidal matrix. 

Now, the first-order part of $\xi^A$ is Lie dragged along $k^a$~\cite{schutz}, so obeys
\be
\dot\xi_a=\xi^b\nabla_b k_a = \frac{1}{2}\theta\xi_a+\sigma_{ab}\xi^b~~~\Rightarrow~~~\dot\xi_A=\frac{1}{2}\theta\xi_A+\sigma_{AB}\xi^B
\ee
In matrix form,
\be\label{sdhsdv}
\dot\x=\S\x~~~\text{where}~~~S_{AB}=\frac{1}{2}\theta\delta_{AB}+\sigma_{AB}
=
\left(
\begin{array}{cc}
\frac{1}{2}\theta+\sigma_1  &   \sigma_2   \\
 \sigma_2 &       \frac{1}{2}\theta-\sigma_1
\end{array}
\right)
\ee
where $\sigma_1=\sigma_{11}=-\sigma_{22},~\sigma_2=\sigma_{12}=\sigma_{21}$. $\S$ is the (linear) optical deformation matrix. Consequently, we have the alternative form for the 2nd order GDE:
\bea\label{dsjhbsjhdbccsbc}
\ddot\xi_A&=&\mathcal{R}_{AB}\xi^B+\mathcal{T}_{ABC}\xi^B\xi^C\,,~~~~\text{where} ~~\\
\mathcal{T}_{ABC}&=&
\frac{1}{2}\dot R_{(BC)A}
-\frac{1}{2}e_A^{~~a}e_{(B}^{~~b}e_{C)}^{~~c}\nabla_c(R_{adbe}k^dk^e)
+\frac{3}{4}{\theta}R_{(BC)A}
-\frac{1}{2}(R_{AD(B}+3R_{(B|AD|})\sigma^D_{~\,C)}
\eea
We shall call this the (optical) distortion tensor.

\subsection{The general solution of the second-order GDE}

We now turn to the solution of \eqref{dsjhbsjhdbccsbc}. We are interested in the case where the geodesic congruence converges at the observer, so that we have initial conditions 
\be
\xi_A(\lambda_o)=0,~~~~ \dot\xi_A(\lambda_o)\neq0\,.
\ee
Since \eqref{dsjhbsjhdbccsbc} arises pertubatively in powers of $\xi$, we can solve it perturbatively by writing 
\be
\xi(\lambda)=\xi_1(\lambda)+\xi_2(\lambda),
\ee
where $\xi_2=\mathcal{O}(\xi_1)^2$ in the usual way.
The linear part is the solution of
\be
\ddot\xi_A=\mathcal{R}_{AB}\xi^B
\ee
with initial conditions at the observer
\be
\xi_A(\lambda_o)=0,~~~~ \dot\xi_A(\lambda_o)=\zeta_A\,.
\ee
The solution may be written in terms of the (linear) Jacobi map, $\mathcal{J}_{AB}$
\be
\xi_A(\lambda)=\mathcal{J}_{AB}(\lambda)\zeta^B
\ee
where $\mathcal{J}_{AB}$ satisfies
\be
\ddot{\mathcal{J}}_{AB}=\mathcal{R}_{AC}\mathcal{J}^C_{~B},~~~\text{with}~~~~\mathcal{J}_{AB}(\lambda_o)=0,~~~\dot{\mathcal{J}}_{AB}(\lambda_o)=-\delta_{AB}\,.
\ee
The Jacobi map takes a deviation vector at the observer and maps it to the deviation vector at the source.
In matrix form we have
\be
\ddot\J=\R \J\,,\label{dsjhbcshjbcs}
\ee
which we shall find convenient later.
Now, the linear $\x_1$ also obeys \eqref{sdhsdv}, because it is Lie dragged along the geodesic congruence~-- this relates the derivatives of $\bm\xi$ to those of $k^a$. Writing this in terms of $\J$ we have
\be\label{sdjhbcsjdhbc}
\dot\J=\S\J\,.
\ee
This implies the optical deformation matrix obeys
\be\label{sdjhbcsjddshbc}
\dot \S+\S\S=\R\,.
\ee
Thus the Jacobi map is found by integrating \eqref{dsjhbcshjbcs}, and is related to the ray bundles expansion and shear through \eqref{sdjhbcsjddshbc}. 

We shall now find the quadratic part of the non-linear GDE, assuming we know the solution to the linear part. Inserting the linear solution in terms of the Jacobi map into \eqref{dsjhbcsbc}, $\xi_2^A$ satisfies
\bea
\ddot\xi_A-\mathcal{R}_{AB}\xi^B&=&F_A\nonumber\\
&=&\left[\mathcal{P}_{ABC}\mathcal{J}^B_{~D}\mathcal{J}^C_{~E}+\mathcal{Q}_{ABC}\mathcal{J}^B_{~D}\dot{\mathcal{J}}^C_{~E}\right]\zeta^D\zeta^E\nonumber\\
&=&\mathcal{T}_{ABC}\mathcal{J}^B_{~D}\mathcal{J}^C_{~E}\zeta^D\zeta^E
\eea
or in matrix form
\be\label{skdjncscn}
\ddot\x-\R\x=\F
\ee
We can solve this by a modified variation of parameters method. For this we need an independent solution to the homogeneous equation $\ddot\x-\R\x=0$ in addition to $\J$. Define a reciprical Jacobi map $\K$:
\be
\ddot\K=\R\K~~~\text{with}~~~\K(\lambda_o)=\I,~~~\dot\K(\lambda_o)=0\,.
\ee
This matrix can be found in terms of $\J$ and $\S$. First, we note that since $\R$ is symmetric, $\R=\R^T$, $\K$ and $\J$ must be related by
\be\label{skjncd}
\dot\J^T\K-\J^T\dot\K=-\I
\ee
which can be checked by differentiating, and by verifying the initial conditions. 
This implies that $\K$ obeys 
\be\label{dsjbc}
\dot\K=\S\K+(\J^T)^{-1}\,.
\ee
We shall assume this solution is known, in addition to $\J$.
Now we suppose there is a solution to \eqref{skdjncscn} of the form
\be
\x=\K \a +\J\b\,,
\ee
where we shall assume 
\be
\K \dot\a +\J\dot\b=0\,.
\ee
Then \eqref{skdjncscn} becomes
\be
\dot\K \dot\a +\dot\J\dot\b=\F\,.
\ee
In matrix form we are then solving
\bea
\left(
\begin{array}{cc}
 \dot\K & \dot\J    \\
 \K &  \J   
\end{array}
\right)
\left(
\begin{array}{cc}
 \dot\a   \\
 \dot\b  
\end{array}
\right)
=\left(
\begin{array}{cc}
 \F  \\
 0  
\end{array}
\right)~~~\Rightarrow~~~
\left(
\begin{array}{cc}
 \a   \\
 \b  
\end{array}
\right)
=\int_{\lambda_o}^\lambda\d\lambda'
\left(
\begin{array}{cc}
 \dot\K & \dot\J    \\
 \K &  \J   
\end{array}
\right)^{-1}\left(
\begin{array}{cc}
 \F  \\
 0  
\end{array}
\right)\,.
\eea
The inverse of a block matrix is given by
\be
\begin{bmatrix}
A & B \\
C & D
\end{bmatrix}^{-1}
=
\begin{bmatrix}
                 (A - BD^{-1}C)^{-1}         & -A^{-1}B(D - CA^{-1}B)^{-1} \\
                 -D^{-1}C(A - BD^{-1}C)^{-1} & (D - CA^{-1}B)^{-1}  
\end{bmatrix}
\ee
which implies
\be
\left(
\begin{array}{cc}
 \dot\K & \dot\J    \\
 \K &  \J   
\end{array}
\right)^{-1}= 
\left(
\begin{array}{cc}
                 ({\dot\K} - {\dot\J}{\J}^{-1}{\K})^{-1}         & -{\dot\K}^{-1}{\dot\J}({\J} - {\K}{\dot\K}^{-1}{\dot\J})^{-1} \\
                 -{\J}^{-1}{\K}({\dot\K} - {\dot\J}{\J}^{-1}{\K})^{-1} & ({\J} - {\K}{\dot\K}^{-1}{\dot\J})^{-1}  
\end{array}
\right)\,.
\ee
We only need the first column of this block matrix. In particular,
\be
({\dot\K} - {\dot\J}{\J}^{-1}{\K})^{-1}=
[{\S\K}+(\J^T)^{-1} - {\S\J}{\J}^{-1}{\K}]^{-1}=\J^T\,.
\ee
Therefore we have 
\be
\left(
\begin{array}{cc}
 \a   \\
 \b  
\end{array}
\right)
=\int_{\lambda_o}^\lambda\d\lambda'
\left(
\begin{array}{cc}
 \J^T \F   \\
 -\J^{-1}\K \J^T \F
\end{array}
\right)\,.
\ee
The second-order solution is then
\be\label{jdshbvshbdshbv}
\x=\int_{\lambda_o}^\lambda\d\lambda'\left[
\K(\lambda)-\J(\lambda)\J^{-1}(\lambda')\K(\lambda')
\right]\J^T(\lambda') \F(\lambda') \,.
\ee

The full solution to the second-order GDE can be written in terms of a linear Jacobi map and a quadratic Hessian as
\be
\xi_A=\mathcal{J}_{AB}\zeta^B+\mathcal{H}_{ABC}\zeta^B\zeta^C
\ee
where the new Hessian part is
\be
\mathcal{H}_{ABC}(\lambda)=\int_{\lambda_o}^\lambda\d\lambda'\big[\mathcal{K}_A^{~\,F}(\lambda)
-\mathcal{J}_A^{~\,D}(\lambda)(\mathcal{J}^{-1})_D^{~~E}(\lambda')
\mathcal{K}_E^{~\,F}(\lambda')
\big]\mathcal{J}_{~~F}^{G}(\lambda')\mathcal{J}^{~~H}_{G}(\lambda')\mathcal{J}^{~~I}_{H}(\lambda')\mathcal{T}_{IBC}(\lambda')\,.
\ee
One can check by differentiating that
\be
\ddot{\mathcal{H}}_{ABC}=\mathcal{R}_A^{~\,D}\mathcal{H}_{DBC}+
\mathcal{T}_{ADE}\mathcal{J}_B^{~\,D}\mathcal{J}_C^{~\,E}\,,
\ee
which gives the differential relation between the Hessian map and the optical distortion tensor. 

\subsection{Extended weak lensing formalism}

We shall now outline the general procedure for finding the Hessian map in terms of the usual weak lensing variables, the convergence and the weak lensing shear (as opposed to the shear of the geodesic ray bundle). 

The Jacobi map can be expanded in terms of a mean distance and an amplification matrix, 
\be
\J=\bar d_A\A
=\bar d_A\left(\begin{array}{cc}1-\kappa-\gamma_1 & \gamma_2 \\\gamma_2 & 1-\kappa+\gamma_1\end{array}\right)=\sqrt{\frac{{\det\J}}{{\det\A}}}\A
\ee 
(We are now assuming that the Jacobi map is symmetric, but extending to the case with rotation is straightforward.) It is normally assumed that this decomposition is around an FLRW background, by identifying 
\be
\bar d_A=\sqrt{\frac{{\det\J}}{{\det\A}}}
\ee
as proportional to the background area distance (proportionality determined via $\dot{\bar d}_A(\lambda_o)=-1$), but really this decomposition of $\J$ is completely general. Within this `weak lensing' interpretation, $\kappa$ is the convergence, and $\gamma_{AB}$ is the trace-free shear, where $\gamma_1=-\gamma_{11}=+\gamma_{22},~\gamma_2=\gamma_{12}=\gamma_{21}$. 

We can expand the reciprical Jacobi map $\K$ in the same manner:
\be
\K=\tilde d_A\widetilde\A
=\tilde d_A\left(\begin{array}{cc}1-\tilde\kappa-\tilde\gamma_1 & \tilde\gamma_2 \\\tilde\gamma_2 & 1-\tilde\kappa+\tilde\gamma_1\end{array}\right)
\ee 
where we shall assume that $\tilde d_A(\lambda)$ is determined from $\J$ in the case when $\kappa=\gamma=0$. From~\eqref{skjncd}, this implies
\be
\frac{\dot{\tilde d}_A}{\tilde d_A}=\frac{\dot{\bar d}_A}{\bar d_A}+\frac{1}{\tilde d_A\bar d_A}
\ee
with $\bar d_A(\lambda_o)=0$. The background area distance and reciprical distance are determined from the background expansion rate $\bar\theta$ via
\be
\dot{\bar d}_A = \frac{1}{2}\bar\theta\bar d_A\,~~~\Rightarrow~~~
\dot{\tilde d}_A = \frac{1}{2}\bar\theta\tilde d_A+\bar d_A^{-1}\,.
\ee
With this, the reciprocal  weak lensing variables $\tilde\kappa,\tilde\gamma_{AB}$ are determined from \eqref{dsjbc}, which becomes the system of linear differential equations:
\bea
\dot{\tilde\kappa}+\left(\frac{1}{\tilde d_A\bar d_A}-\frac{1}{2}\Delta\theta\right)\tilde\kappa-\sigma_1\tilde\gamma_1+\sigma_2\tilde\gamma_2
&=&-\frac{1}{2}\Delta\theta+\frac{1-(1-\kappa)\mu}{\tilde d_A\bar d_A}\,,\\
\dot{\tilde\gamma}_1+\left(\frac{1}{\tilde d_A\bar d_A}-\frac{1}{2}\Delta\theta\right)\tilde\gamma_1-\sigma_1\tilde\kappa&=&-\sigma_1-\frac{\gamma_1\mu}{\tilde d_A\bar d_A}\,,\\
\dot{\tilde\gamma}_2+\left(\frac{1}{\tilde d_A\bar d_A}-\frac{1}{2}\Delta\theta-\sigma_1\right)\tilde\gamma_2+\sigma_2\tilde\kappa-\sigma_2\tilde\gamma_1&=&\sigma_2-\frac{\gamma_2\mu}{\tilde d_A\bar d_A}\,.
\eea
We have defined $\Delta\theta=\theta-\bar\theta$ for convenience~-- but there are no approximations made. 

The weal lensing convergence and shear are found in terms of the Sachs optical scalars from~\eqref{sdjhbcsjdhbc}, which become
\bea
\dot\kappa-\frac{1}{2}\Delta\theta\kappa-\sigma_1\gamma_1+\sigma_2\gamma_2&=& -\frac{1}{2}\Delta\theta\,,\\
\dot\gamma_1-\frac{1}{2}\Delta\theta\gamma_1-\sigma_1\kappa&=&-\sigma_1\,,\\
\dot\gamma_2-\left(\frac{1}{2}\Delta\theta+\sigma_1\right)\gamma_2
+\sigma_2\kappa-\sigma_2\gamma_1&=&\sigma_2\,.
\eea

The Sachs optical scalars are found from~\eqref{sdjhbcsjddshbc}.  The trace of \eqref{sdjhbcsjddshbc} is
\be
\dot\theta+\frac{1}{2}\theta^2+2(\sigma_1^2+\sigma_2^2)=-R_{ab}k^ak^b,
\ee
while the trace-free part becomes
\be
\dot\sigma_{AB}+\theta\sigma_{AB}=C_{ABcd}k^ck^d\,,
\ee
which are just the Sachs optical equations as derived from the Ricci identities, but now derived from the GDE combined with the fact that the linear deviation vector is Lie dragged along the congruence.

The Hessian of the transformation is an integrated projection of the optical distortion tensor,
\be
\mathcal{H}_{ABC}(\lambda)=\mathcal{P}_A^{~\,D}[\mathcal{T}_{DBC}](\lambda)~~~\text{or}~~~{\bm{\mathcal{H}}}_{BC}(\lambda)=\P[{\bm{\mathcal{T}}}_{BC}](\lambda)\,,\label{jnvdknv}
\ee
where we have used a matrix notation for the first index of the Hessian and distortion tensor. The projection operator $\P$ is 
\bea\label{dfkjnvskjdfs}
\P&=&\int_{\lambda_o}^\lambda\d\lambda'\left[
\K(\lambda)-\J(\lambda)\J^{-1}(\lambda')\K(\lambda')
\right]\J^T(\lambda')\J^2(\lambda') \nonumber\\
&=&\int_{\lambda_o}^\lambda\d\lambda'\bar d_A(\lambda')^2\left[
\tilde d_A(\lambda)\bar d_A(\lambda') \widetilde{\A}(\lambda)\widetilde{\A}^{-1}(\lambda')-
\bar d_A(\lambda)\tilde d_A(\lambda') {\A}(\lambda){\A}^{-1}(\lambda')
\right]\widetilde{\A}(\lambda'){\A}^3(\lambda')
\eea

We now have everything in place to construct the Hessian of the lensing map. The strategy is to construct $\K$ and $\J$ as follows:
\begin{enumerate}
\item Solve for $\bar\theta$ and then $\bar d_A$ and $\tilde d_A$.
\item Find the Sachs optical scalars $\theta$ and $\sigma_{AB}$.
\item Find the components of the amplification matrix, $\kappa$ and $\gamma_{AB}$.
\item Find the components of the reciprical amplification matrix, $\tilde \kappa$ and $\tilde\gamma_{AB}$.
\item Calculate the components of the optical distortion tensor~\eqref{kjscdjcnkscn}.
\item Construct the integrated projection operator $\P$~\eqref{dfkjnvskjdfs}.
\item Integrate the distortion tensor against the projection operator  to calculate the Hessian from \eqref{jnvdknv}.

\end{enumerate}

\subsection{Influence of the Ricci vs Weyl curvature}

We can expand the optical tidal and distortion tensors in terms of the general matter variables and the Weyl curvature tensor, which we shall split into its invariant electric and magnetic parts.
The Ricci tensor and scalar are related to the energy momentum tensor via
\be
R_{ab}=T_{ab}-\sfrac{1}{2}Tg_{ab}+\Lambda g_{ab}
\ee
The stress energy tensor has the invariant decomposition given the observer velovity $u^a$:
\be
T_{ab}=\rho u_a u_b+p h_{ab}+2q_{(a} u_{b)}+\pi_{ab}\,.
\ee
Then the fluid part of the Riemann tensor is
\bea
(R_{\text{fluid}})^{ab}{}_{cd} & = & \sfrac{2}{3}\,(\rho+3p-2\Lambda)\,u^{[a}\,
u_{[c}\, h^{b]}{}_{d]} + \sfrac{2}{3}\,(\rho+\Lambda)\,h^{a}{}_{[c}
\,h^{b}{}_{d]} \nonumber\\&& -\,2\,u^{[a}\,h^{b]}{}_{[c}\,q_{d]}
- 2\,u_{[c}\,h^{[a}{}_{d]}\,q^{b]}
- 2\,u^{[a}\,u_{[c}\,\pi^{b]}{}_{d]} 
+ 2\,h^{[a}{}_{[c}\,\pi^{b]}{}_{d]} \,.
\eea
The Weyl tensor can be decomposed into an electric and magnetic part:
\bea
E_{ab} = C_{acbd}\,u^c\,u^d
~~~~~~
H_{ab} = {\sfrac12}\,\varepsilon_{ade}\,C^{de}{}_{bc}\,u^{c}\,,
\eea
so that the Weyl tensor is expanded as 
\be
C^{ab}{}_{cd}  =  4\,u^{[a}\,u_{[c}\,E^{b]}{}_{d]} 
+ 4\,h^{[a}{}_{[c}\,E^{b]}{}_{d]}+ 2\,\varepsilon^{abe}\,u_{[c}\,H_{d]e} 
+ 2\,\varepsilon_{cde}\,u^{[a}\,H^{b]e} \ . \nonumber
\ee
The projections of these tensors we require are:
\bea
R_{ab}k^a k^b&=& \kappa E^2\left(\rho+p+2q\p+\pi\p\right)\,,\\
R_A^{~\,b}k_b&=& -\kappa E (q_A+{\pi\p}_A)\,,\\
C_{AcBd}k^ck^d &=& E^2\left(\delta_{AB}E\p+2E_{AB}-2\varepsilon_{C(A}H_{B)}^{~~C}
\right)\,,\\
C_{AdBC}k^d&=&E\left(2\delta_{A[C}{E\p}_{B]}+\varepsilon_{BC}{H\p}_A
\right)
\eea
so that the projected parts of Riemann which appear in the GDE are
\bea
E^{-2}\mathcal{R}_{AB}&=&-\frac{1}{2}\delta_{AB}\left(\rho+p+2q\p+\pi\p+2E\p\right)-2E_{AB}+2\varepsilon_{C(A}H_{B)}^{~~C}\,,\\
E^{-1}{R}_{ABC}&=&- \delta_{A[B}(q_{C]}+{\pi\p}_{C]}+2{E\p}_{C]})+\varepsilon_{BC}{H\p}_A\,,\\
E^{-1}{\dot R}_{ABC}&=&-\delta_{A[B}(\dot q_{C]}+{\dot{\pi}}_{\text{\tiny$\|$}C]}+2{\dot E}_{\text{\tiny$\|$}C]})+\varepsilon_{BC}{\dot H}_{\text{\tiny$\|$}A}\nonumber\\&&
+E\left(\frac{1}{3}\Theta -  A\p+ \Sigma\p\right)\left[
 \delta_{A[B}(q_{C]}+{\pi\p}_{C]}+2{E\p}_{C]})-\varepsilon_{BC}{H\p}_A
\right]\,.
\eea
Here we are using compressed notation whereby
\be
{\dot E}_{\text{\tiny$\|$}A}=\frac{D}{\d\lambda}(n^bE_{bA}) = 
E\left[E_{\text{\tiny$\|$}A}(A\p-\Sigma\p)+(-A^b+\Sigma\p^{~b}+\Omega\p^{~b})E_{bA}
\right]+n^b\dot E_{bA}
\,.
\ee
We also require the projected derivative of the optical tidal tensor:
\bea
E^{-2}e_A^{~~a}e_B^{~~b}e_C^{~~c}\nabla_a(R_{bdce}k^dk^e)
&=& \frac{1}{2}\delta_{BC}
\left[\nabla_A(\rho+p+2q\p+\pi\p+2E\p)+2({\Sigma\p}_A+{\Omega\p}_A)(\rho+p+2q\p+\pi\p+2E\p)
\right]\nonumber\\&&
-2\left(\frac{1}{3}\Theta\delta_{A(C}+\Sigma_{A(C}-\Omega_{A(C}\right)
(2{E\p}_{B)}-{{\overline H}\p}_{B)})
\nonumber\\&&
+\frac{1}{E}\left(\frac{1}{2}\theta\delta_{A(C}+\sigma_{A(C}\right)
\left({q\p}_{B)}+{\pi\p}_{B)}+2{E\p}_{B)}\right)
\nonumber\\&&
+4({\Sigma\p}_A+{\Omega\p}_A)\left(E_{BC}+\overline H_{BC}\right)
+
2e_A^{~~a}e_B^{~~b}e_C^{~~c}\left(\nabla_aE_{bc}-\varepsilon_{d(c}\nabla_{|a|}H_{b)}^{~~d}\right)\,.
\eea
We have used an over-bar to denote the reverse parity in the screen space:
\be
{\overline{X}}_A=\varepsilon_{AB}X^B~~~\text{and}~~~{\overline{X}}_{AB}=\varepsilon_{C(A}X_{B)}^{~~C}\,.
\ee
Consequently, the optical distortion tensor may be written as sum of parts induced by Ricci curvature $\mathcal{T}_{ABC}^{{}^\text{Ricci}}$, and by non-local Weyl curvature,  $\mathcal{T}_{ABC}^{{}^\text{Weyl}}$, where:
\bea
E^{-1}\mathcal{T}^{{}^\text{Ricci}}_{ABC}&=&
\frac{1}{2}\delta_{A(B}\bigg[-\frac{E}{2}\left(\nabla_{C)}+2{\Sigma\p}_{C)}+2{\Omega\p}_{C)}\right)(\rho+p+2q\p+\pi\p)
+\frac{1}{2}\left({{\dot{q}}\p}{}_{C)}+{\dot{\pi}\p}{}_{C)}\right)
\nonumber\\&&
-{E}\left(\frac{1}{3}\Theta-A\p+\Sigma\p-\frac{3\theta}{4E}\right)\left({q\p}{}_{C)}+{\pi\p}{}_{C)}\right)
+\sigma_{C)}^{~~D}\left({q\p}{}_{D}+{\pi\p}{}_{D}\right)
\bigg]
\nonumber\\&&
+\delta_{BC}\bigg[-\frac{1}{4}\left({{\dot{q}}\p}{}_{A}+{\dot{\pi}\p}{}_{A}\right)+
\frac{E}{4}\left(\frac{1}{3}\Theta-A\p+\Sigma\p-\frac{3\theta}{4E}\right)\left({q\p}_{A}+{\pi\p}_{A}\right)
\bigg]
\nonumber\\&&
+\frac{1}{4}\sigma_{A(B}\left({q\p}_{C)}+{\pi\p}_{C)}\right)
-\frac{3}{4}\sigma_{BC}\left({q\p}_{A}+{\pi\p}_{A}\right)
\nonumber\\&&
-\frac{1}{4}\left(\frac{1}{2}\theta\delta_{C(B}+\sigma_{C(B}\right)
\left({q\p}_{A)}+{\pi\p}_{A)}\right)
-\frac{1}{4}\left(\frac{1}{2}\theta\delta_{B(C}+\sigma_{B(C}\right)
\left({q\p}_{A)}+{\pi\p}_{A)}\right)
\eea
\bea
E^{-1}\mathcal{T}^{{}^\text{Weyl}}_{ABC}&=&\frac{1}{2}\delta_{A(B}\bigg[{\dot E\p}{}_{C)}-E
\left(\nabla_{C)}+2{\Sigma\p}_{C)}+2{\Omega\p}_{C)}\right)E\p
-{E}\left(\frac{1}{3}\Theta-A\p+\Sigma\p-\frac{3\theta}{2E} \right){E\p}_{C)}+2\sigma_{C)}^{~~D}{E\p}_D
\bigg]
\nonumber\\&&
+\frac{1}{2}\delta_{BC}\bigg[-\dot E\p{}_A
+{E}\left(\frac{1}{3}\Theta-A\p+\Sigma\p-\frac{3\theta}{2E} \right){E\p}_{A}
\bigg]
+\frac{1}{2}\varepsilon_{A(B}\left[
-{\dot H\p}{}_{C)}
+{E}\left(\frac{1}{3}\Theta-A\p+\Sigma\p-\frac{3\theta}{2E} \right){H\p}_{C)}\right]
\nonumber\\&&
\left[\frac{1}{3}\Theta\delta_{(C(B)}+\Sigma_{(C(B)}-\Omega_{(C(B)}-\frac{\theta}{4E}\delta_{(C(B)}-\frac{1}{2E}\sigma_{(C(B)}\right]{E\p}_{A)}
+\left[\frac{1}{3}\Theta\delta_{(C(B)}+\Sigma_{(C(B)}-\Omega_{(C(B)}\right]{\overline{H}\p}_{A)}
\nonumber\\&&
+\frac{1}{2}\left(\sigma_{A(B}{E\p}_{C)}-3\sigma_{BC}{E\p}_A\right)
-\frac{1}{2}\overline{\sigma}_{BC}{H\p}_A+\frac{3}{2}\varepsilon_{DA}\sigma^D_{~(C}{H\p}_{B)}
\nonumber\\&&
-2E\left({\Sigma\p}_{(B}+{\Omega\p}_{(B}\right)\left({E}_{C)A}+{{\overline{H}}}_{C)A}\right)
+Ee_A^{~~a}e_{(B}^{~~\,b}e_{C)}^{~\,~c}\nabla_{b}\left(-E_{ca}+\varepsilon_{d(c}{H}_{a)}^{~d}\right)
\label{kjscdjcnkscn}
\,.
\eea
In the last term in \eqref{kjscdjcnkscn} we have refrained from using a separate notation for the fully projected derivative to avoid possible confusion. The double symmetrisation means $2(C(B)A)=C(BA)+B(CA)$.

\subsection{The effect of the invariant parts of the Hessian}

\begin{figure}[t]
\begin{center}
\includegraphics[width=\textwidth]{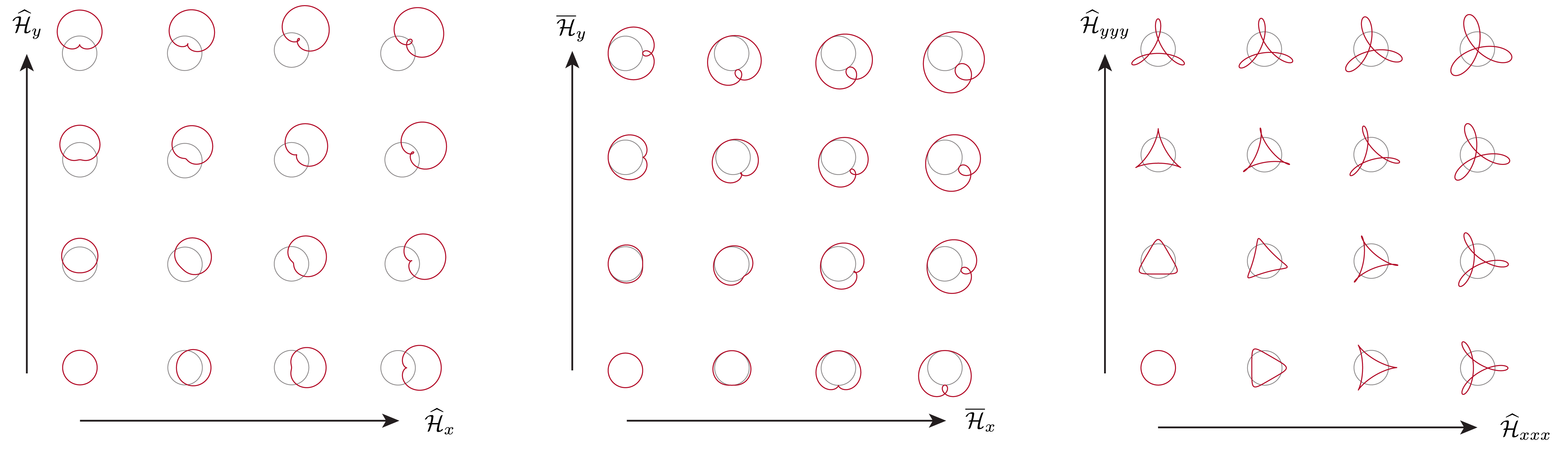}
\caption{The 6 independent degrees of freedom of the Hessian, as defined by separating the anti-symmetric and trace-free parts of the tensor. This is the action on a unit circle. }
\label{sajdncksajckjdnajskc}
\end{center}
\end{figure}

In general, a rank~3 tensor can be invariantly decomposed into antisymmetric, and symmetric trace-free parts~\cite{Maartens:1996ch}. In the case of $\HH_{ABC}$, which is a rank-3 tensor in 2 dimensions satisfying $\HH_{ABC}=\HH_{A(BC)}$, the invariant parts are: $\widehat\HH_{ABC}$, which is totally trace-free, and 2 vectors
$\HA_A$ and $\HB_A$, where
\bea
\HA_A&=& \frac{3}{4}\delta^{BC}\HH_{ABC}\\
\HB_A&=& -\frac{2}{3}\varepsilon_A^{~~B}\delta^{CD}\left(\HH_{[BC]D}+\HH_{[B|D|C]}
\right)\,.
\eea
Then the Hessian may be written
\be
\HH_{ABC}=\HA_{(A}\delta_{BC)}+\varepsilon_{A(B}\HB_{C)}+\widehat\HH_{ABC}\,.
\ee
This gives a total of 6 degrees of freedom~-- 2 in each component. The number of indices indicates the spin level of the invariant parts~-- there are two spin 1 parts, $\HA_{A}$ and $\HB_{A}$ and one spin 3 part~$\widehat\HH_{ABC}$. These correspond to $F$ flexion, twist and $G$-flexion in the language of~\cite{Bacon:2008zj}. Because there are traces in each term in $\mathcal{T}^{{}^\text{Ricci}}_{ABC}$, Ricci curvature can only induce spin 1 degrees of freedom, while (non-local) Weyl curvature can produce all 3, and is the only source of $\widehat\HH_{ABC}$.

Given an the  angle at the observer between neighbouring geodesics, $\zeta^A$, the Jacobi and Hessian maps transform this at the source into
\bea
\xi_A&=&\mathcal{J}_{AB}\zeta^B+\mathcal{H}_{ABC}\zeta^B\zeta^C\\
&=&\bar d (1-\kappa)\zeta_{A}+\bar d\gamma_{AB}\zeta^B
+\HA_{(A}\delta_{BC)}\zeta^B\zeta^C+\varepsilon_{AB}\HB_{C}\zeta^B\zeta^C+\widehat\HH_{ABC}\zeta^B\zeta^C\,.
\eea
From this expression it is clear the action of the Jacobi map: the trace induces a uniform change in area of a source, while the shear induces two area preserving elliptical distortions at 45$^{\circ}$ to each other. To illustrate how $\widehat\HH_{ABC}$, $\HA_A$ and $\HB_A$ change the image of a source, let us examine each in turn, and their action on a circular image. First, $\HA_A$:
\be
\HA_{(A}\delta_{BC)}\zeta^B\zeta^C=\frac{1}{3}\zeta^2\left(\HA_A+2\HA_B\hat\zeta^B\,\hat\zeta_A\right)=\frac{1}{3}\zeta^2\left(
\begin{array}{c}
\HA_x[2+\cos2\theta]+\HA_y\sin2\theta\\
\HA_x\sin2\theta+\HA_y[2-\cos2\theta]
\end{array}
\right)
\ee
where $\zeta_A=|\zeta|\hat\zeta_A=|\zeta|(\cos\theta,\sin\theta)$ in Cartesian coordinates in the plane. This produces a shift in the position of the image, and a change in area. There is a compression along the direction of $\HA_{A}$, leading to a cusp for large distortions. Next, 
\be
\varepsilon_{AB}\HB_{C}\zeta^B\zeta^C = \zeta^2\varepsilon_{AB}\hat\zeta^B\,\, \HB_{C}\hat\zeta^C=\frac{1}{2}\zeta^2\left(
\begin{array}{c}
\HB_x\sin2\theta+\HB_y[1-\cos2\theta]\\
-\HB_x[1+\cos2\theta]-\HB_y\sin2\theta
\end{array}
\right)\,.
\ee
This produces a sagging like distortion, perpendicular to $\HB_{A}$, with the normal modes at 90$^\circ$. Finally the purely trace-free part of the Hessian also has two independent degrees of freedom, $\widehat\HH_{xxx}$ and $\widehat\HH_{yyy}$ (all other components are proportional to one or the other of these since $\widehat\HH_{ABC}$ is trace-free):
\be
\widehat\HH_{ABC}\zeta^B\zeta^C=\zeta^2\left(
\begin{array}{c}
\widehat\HH_{xxx}\cos2\theta-\widehat\HH_{yyy}\sin2\theta\\
-\widehat\HH_{xxx}\sin2\theta-\widehat\HH_{yyy}\cos2\theta
\end{array}
\right)\,.
\ee
This produces a triangular distortion at low amplitude, with polarisations at 30$^\circ$.
These curves are all trochoids of higher complexity than the circle and ellipse which appear at linear order~-- see Fig.~\ref{sajdncksajckjdnajskc}. (A trochoid is the locus of points traced out by a point attached at some distance from a circle, which itself is rolling around a larger circle.) For large amplitudes, we can see cusps appearing representing the formation of caustics.

\section{Perturbations about an FLRW background}\label{kjdsskjdvbkdjvbs}

We shall now linearise the Hessian around a flat FLRW model.  We shall write our perturbations with respect to the Poisson gauge, where 
\begin{eqnarray}\label{metric}
\d  s^2 &=&
 a^2\big[-(1 + 2\Phi )\d \eta^2 
 + (1-2 \Psi)\gamma_{i j}\d x^{i}\d x^{j}\big]\,, 
\end{eqnarray}
where $\gamma_{ij} = \delta_{ij}$ if $x^i$ are cartesian ($i,j,k,\ldots$ denote spatial indices, usually on the conformal Minkowski background when they are raised and lowered with $\gamma_{ij}$). 
The observer 4-velocity $ u^a$ is perturbed as
\bea
 u^0 &=&\frac{1}{a}(1 - \Phi ),
~~~
 u^i=\frac{1}{a}\nabla^iv,~~~u_i=a\nabla_iv
\eea
where $v$ is the first-order scalar velocity potential which obeys
 \be
v = -\frac{2a}{ 3H_0^2\Omega_m}\big(\Psi'+\mathcal{H}\Phi \big)=\frac{\Psi'+\mathcal{H}\Phi}{\HH'-\HH^2}\,, \label{oi}
 \ee
 where $\mathcal{H}=a'/a$ is the conformal Hubble rate, and we also use $H=\Theta/3=\mathcal{H}/a$ in the background, and the background density is $\bar\rho = 2(\HH^2-\HH')/a^2$.
We also have for the time and radial parts of the vector $n^a$,
\be
\delta n^0=\delta u^\chi = \frac{1}{a}\partial_\chi v,~~~\delta n^\chi = \Psi\,,
\ee 
and similarly for $k^a=E(u^a-n^a)$,
\bea
\delta k^0&=&(1+z)\delta E-(1+z)^2(\Phi+\partial_\chi v)\,,\\
\delta k^\chi&=&-(1+z)\delta E-(1+z)^2(\Psi-\partial_\chi v)\label{ksdjvndjfbd}
\eea
where the perturbed energy is given by (enforcing that the time component of the photon 4-vector is unperturbed at the observer)
\be
\delta E = \delta z+(1+z)(\Phi_o +\partial_\chi v_o)
\ee
with the redshift perturbation given by
\be
\frac{\delta z}{1+z} = (\partial_\chi v - \Phi)\big|_0^\chi
-\int_0^\chi\!\d\chi(\Phi'+\Psi')\,.
\ee
Note that $g_{ab}k^ak^b=0$ implies $\delta k^0+\delta k^\chi=-(1+z)^2(\Phi+\Psi)$, giving \eqref{ksdjvndjfbd}.
 
The density contrast $\delta$ is given by
\be
a^2\rho\delta=2\nabla^2\Psi-6\mathcal{H}(\Psi'+\mathcal{H}\Phi)\,.
\ee
For a perfect fluid in GR we have $\Psi=\Phi$. However, we will keep these potentials separate for generality. In that case we have an effective pressure and anisotropic pressure perturbation given by
\bea
a^2\delta p &=& 2\Psi''+2\mathcal{H}(2\Psi'+\Phi')+2(2\mathcal{H}'+\mathcal{H}^2)\Phi-\frac{2}{3}\nabla^2(\Psi-\Phi)\,,\\
\pi_{ij}&=&\nabla_{\langle i}\nabla_{ j\>}(\Psi-\Phi)\,.
\eea
In the LCDM case in GR, the first leads to the Bardeen equation
$\Phi''+3\mathcal{H}\Phi'+a^2\Lambda\Phi=0\,.$

We also require the matter shear and electric Weyl tensor. These are
\bea
\Sigma_{ij}&=&a\nabla_{\langle i}\nabla_{ j\>}v=a
\nabla_{ i}\nabla_{ j}v-\frac{a}{3}\gamma_{ij}\nabla^2 v\,,\\
E_{ij}&=&\frac{1}{2}\nabla_{\langle i}\nabla_{ j\>}(\Psi+\Phi)=\frac{1}{2}\nabla_{ i}\nabla_{ j}(\Psi+\Phi)-\frac{1}{6}\gamma_{ij}\nabla^2(\Psi+\Phi)\,.
\eea
 
We denote the radial co-moving coordinate as $\chi$, where 
\be
a\,n^i\nabla_i=\frac{\partial}{\partial\chi}=  \frac{\d}{\d\chi}+
\frac{\partial}{\partial\eta}
\ee
in the background. Because $\mathcal{T}_{ABC}$ is first-order we do not need the perturbation of this. The derivative $\d/\d\chi$ is along the null geodesic, so that $\chi$ is a conformal affine parameter from the observer to the source obeying $\d\lambda/\d\chi=-a^2$. Together with this we have the spatial Laplacian,
\be
\nabla_i\nabla^i=h^{ab}\nabla_a\nabla_b = a^{-2} \gamma^{ij}\nabla_i\nabla_j=a^{-2}\nabla^2\,,
\ee
with the notation such that $\nabla^2$ is only used to denote the co-moving spatial Laplacian (i.e., the Laplacian on a Minkowski background). This is expanded using $N_{ab}$ into a co-moving screen Laplacian $\nabla_\o^2$ and radial parts, as
\bea
\nabla^2&=&\nabla_\o^2+\partial_\chi^2+\frac{2}{\chi}\partial_\chi\nonumber\\
&=&\nabla_\o^2+{\d^2~\over \d\chi^2}+2{\d\over \d\chi}\partial_\eta+\partial_\eta^2+{2\over \chi}\left(\partial_\eta+{\d\over \d\chi}\right).
\eea
We also require a comoving tetrad in the screen-space (i.e., with the scale-factor factored out), which is the natural tetrad for the observer. We define these as 
\be
a\bm e_I=a\{\bm n,\bm e_A\}=\hat{\bm e}_I=\{{\hat{\bm e}_\chi,\hat{\bm e}_A}\}~~~\text{where}~~~I={\chi,A},~~~\text{and}~~~A={\vartheta,\varphi}\,.
\ee
 With respect to a standard Cartesian basis $\bm i, \bm j,\bm k$, we have
 \bea
 \hat{\bm e}_\chi&=&\sin\vartheta\cos\varphi\,\bm i+\sin\vartheta\sin\varphi\,\bm j+\cos\vartheta\,\bm k\,,\\
 \hat{\bm e}_\vartheta&=&\cos\vartheta\cos\varphi\,\bm i+\cos\vartheta\sin\varphi\,\bm j-\sin\vartheta\,\bm k\,,\\
 \hat{\bm e}_\varphi&=&-\sin\varphi\,\bm i+\cos\varphi\,\bm j\,,
 \eea
  and
\be
\bm\nabla=\hat{\bm e}_\chi\partial_\chi+\frac{\hat{\bm e}_\vartheta}{\chi}\partial_\vartheta+\frac{\hat{\bm e}_\varphi}{\chi\sin\vartheta}\partial_\varphi\,.
\ee
The Ricci rotation coefficients for this tetrad are defined by
\be
\nabla_I\hat{\bm e}_J=(\hat{\bm e}_I\cdot\bm\nabla)\hat{\bm e}_J=\Gamma^K_{~~JI}\hat{\bm e}_K~~~\Rightarrow~~~\left\{
\begin{array}{l}
\Gamma^{A}_{~~\chi A }=-\Gamma^{\chi}_{~AA}=\displaystyle\frac{1}{\chi}~~~\Rightarrow~~~\Gamma^\chi_{~AB}=-\frac{1}{\chi}\delta_{AB}\\
\Gamma^{\vartheta}_{~\varphi\varphi}=-\Gamma^{\varphi}_{~\vartheta\varphi}=-\displaystyle\frac{\cot\vartheta}{\chi}~~~\Rightarrow~~~\Gamma_{ABC}=2\varepsilon_{AB}\delta_C^{~\varphi}\,\displaystyle\frac{\cot\vartheta}{\chi}
\end{array}
\right.
\ee
(no sum implied in the expressions on the right). The tetrad alternating tensor is $\epsilon_{AB}=\delta_A^{~[\vartheta}\delta_B^{~\varphi]}$. We are using notation where $\nabla_I=\hat{\bm e}_I^b\nabla_b$, a derivative which we only use on scalars or tensors projected onto the tetrad basis. In this basis the 3-d Laplacian is
\be
\nabla_I\nabla^I=\delta^{IJ}(\hat{\bm e}_I\cdot\bm \nabla)(\hat{\bm e}_J\cdot\bm \nabla)-\delta^{IJ}(\hat{\bm e}_I\cdot\nabla\hat{\bm e}_J)\cdot\bm \nabla = \partial_\chi^2 +\nabla_A\nabla^A
\ee
which implies that the 2-D Laplacian is,
\be
\nabla_\o^2=\nabla_A\nabla^A-\frac{2}{\chi}\partial_\chi
=\nabla_A\nabla^A-\frac{2}{\chi}\left(\frac{\d}{\d\chi}+\partial_\eta\right)
\,,
\ee
which in turn implies the 3d Laplacian is
\be
\nabla^2=\nabla_A\nabla^A+\left(\frac{\d}{\d\chi}+\partial_\eta\right)^2\,.
\ee
Radial and angular derivatives commute when acting on scalars as 
\bea
\nabla_A\partial_\chi=\partial_\chi\nabla_A+\frac{1}{\chi}\nabla_A&=&\frac{\d}{\d\chi}\nabla_A+\frac{1}{\chi}\nabla_A+\nabla_A\partial_\eta\,,\\
\nabla_A\frac{\d}{\d\chi} &=& \frac{\d}{\d\chi}\nabla_A+\frac{1}{\chi}\nabla_A\,,\\
\nabla_A\frac{\d^2}{\d\chi^2}  &=& \left(\frac{\d}{\d\chi} + \frac{2}{\chi}\right)\frac{\d}{\d\chi}\nabla_A\,.
\eea
The integral version of the second is found from $\chi\nabla_A\d/\d\chi f=\d/\d\chi(\chi\nabla_A f)$ as
\be
\chi\nabla_A\int_0^\chi\d\chi' = \int_0^\chi\d\chi'\chi'\nabla_A\,.
\ee
Projecting derivatives on scalars we have:
\bea
\hat e_A^{~~a}\hat e_B^{~~b}\nabla_a\nabla_b&=&\nabla_A\nabla_B+\frac{1}{\chi}\delta_{BA}\partial_\chi-\Gamma^C_{~~AB}\nabla_C\,,\\
\hat e_A^{~~a}\hat e_B^{~~b}\hat e_C^{~~c}\nabla_a\nabla_b\nabla_c&=&
\nabla_A\nabla_B\nabla_C+\frac{3}{\chi}\delta_{(AB}\nabla_{C)}\partial_\chi
-\Gamma^D_{~~BA}\nabla_{C}\nabla_D
-2\Gamma^D_{~~C(A}\nabla_{B)}\nabla_D
\nonumber\\&&
+
\left[2\Gamma^D_{~~E(C}\Gamma^E_{~~B)A}-\nabla_{A}\Gamma^D_{~~CB}\right]
\nabla_D
-\frac{2}{\chi}\Gamma_{(B|A|C)}\partial_\chi-\frac{2}{\chi^2}\delta_{A(B}\nabla_{C)}
\eea


Relative to the matter frame, the  matter content is dust. For full generality we shall keep the effective pressure components non-zero so that our results can be used in the case of modified gravity. To linear order the magnetic Weyl tensor is zero for scalar modes. So, the distortion tensor becomes, relative to the co-moving matter frame
\bea
\mathcal{T}^{{}^\text{Ricci}}_{ABC}&=&
-\frac{1}{4}{(1+z)^2}\delta_{A(B}\left(\nabla_{C)}\rho+2\bar\rho\,{\Sigma\p}_{C)}\right)\nonumber\\
&&-\frac{1}{4}(1+z)\delta_{A(B}\left[(1+z)\nabla_{C)}(p+\pi\p)-{\dot{\pi}\p}{}_{C)}
+\left(-\frac{3}{2}\theta+(1+z)H\right){{\pi}\p}{}_{C)}
\right]\nonumber\\&&
-\frac{1}{4}\delta_{BC}\left[{\dot{\pi}\p}{}_{A}
-\left(-\frac{3}{2}\theta+(1+z)H\right){{\pi}\p}{}_{A}
\right]
\\
\mathcal{T}^{{}^\text{Weyl}}_{ABC}&=&\frac{1}{2}(1+z)\delta_{A(B}\left\{-(1+z)\nabla_{C)}E\p+{\dot E\p}{}_{C)}
+\frac{5}{4}\theta{E\p}_{C)}
\right\}\nonumber\\&&
+\frac{1}{2}(1+z)\left[-{\dot E\p}{}_{A}+\left(-\frac{7}{4}\theta+2(1+z)H\right){E\p}_{A}
\right]\delta_{BC}
-(1+z)^2e_A^{~~a}e_{(B}^{~~\,b}e_{C)}^{~\,~c}\nabla_{b}E_{ca}\,.
\eea

The projected parts of $E_{ab}$ and $\Sigma_{ab}$ we require are:
\bea
2a^2E\p=2\hat e_\chi^{~a}\hat e_\chi^{~b}E_{ab}&=&\left[\partial_\chi^2-\frac{1}{3}\nabla^2\right](\Phi+\Psi)
=\left[-\frac{1}{3}\nabla_A\nabla^A+\frac{2}{3}\left(\frac{\d}{\d\chi}+\partial_\eta\right)^2\right](\Phi+\Psi)
\,,\\
2a^2{E\p}_A=2\hat e_A^{~a}\hat e_\chi^{~b}E_{ab}&=&\left[\frac{\d}{\d\chi}\nabla_A+\nabla_A\partial_\eta\right](\Phi+\Psi)\,,\\
%
2a^3 e_{A}^{~\,~a}e_{(B}^{~~b}e_{C)}^{~~\,b}\nabla_{b}E_{ca}&=&
\bigg[\nabla_{(B}\nabla_{C)}\nabla_A-\frac{1}{3}\delta_{A(C}\nabla_{B)}\nabla^2+\frac{3}{\chi}\delta_{(BC}\nabla_{A)}\partial_\chi
-\Gamma^D_{~~(BC)}\nabla_{A}\nabla_D
\nonumber\\&&
-2\Gamma^D_{~~A(B}\nabla_{C)}\nabla_D
+
\left[\Gamma^D_{~~EA}\Gamma^E_{~~(BC)}+\Gamma^D_{~~E(C}\Gamma^E_{~~|B|A)}-\nabla_{(B}\Gamma^D_{~~|A|C)}\right]
\nabla_D
\nonumber\\&&
-\frac{1}{\chi}\Gamma_{A(BC)}\partial_\chi
-\frac{1}{\chi^2}\delta_{BC}\nabla_{A}
-\frac{1}{\chi^2}\delta_{A(B}\nabla_{C)}\bigg](\Phi+\Psi)\,,\\
2a^4{\dot E\p}{}_{A}&=&\bigg[2\mathcal{H}\left(\frac{\d}{\d\chi}\nabla_A+\nabla_A\partial_\eta\right)-\frac{\d^2}{\d\chi^2}\nabla_A-\frac{\d}{\d\chi}\nabla_A\partial_\eta\bigg](\Phi+\Psi)
\\
a^3\rho{\Sigma\p}_A&=&
-2\left(\frac{\d}{\d\chi}+\partial_\eta\right)\nabla_A\left(\Psi'+\mathcal{H}\Phi\right)
\,,\\
a^2\pi\p&=&\left[\partial_\chi^2-\frac{1}{3}\nabla^2\right](\Psi-\Phi)
=
\left[-\frac{1}{3}\nabla_A\nabla^A+\frac{2}{3}\left(\frac{\d}{\d\chi}+\partial_\eta\right)^2\right](\Psi-\Phi)
\,,\\
a^2{\pi\p}_A&=&\left[\frac{\d}{\d\chi}\nabla_A+\nabla_A\partial_\eta\right](\Psi-\Phi)\,,
\eea
Note the notation here: on the left hand side we are using the usual tetrad $\bm e_I$, while on the right the indices are with respect to $\hat{\bm e}_I$. Really we should introduce new notation for objects in the tetrad $\hat{\bm e}_I$, but we don't need to: if an object is a 1+3 covariant tensor used above we're using $\bm e_I$; if there's $\Phi$'s then the indices represent the comoving tetrad $\hat{\bm e}_I$. We also require
\bea
\frac{a^2}{2}\nabla_A\rho &=& \nabla_A\nabla_B\nabla^B\Psi
+\left[\frac{\d^2}{\d\chi^2}+\partial_\eta^2 + 2\left(\partial_\eta +\frac{1}{\chi}\right)\frac{\d}{\d\chi}+\left(\frac{2}{\chi}-3\HH\right)\partial_\eta
\right]\nabla_A\Psi - 3\HH^2\nabla_A\Phi\,.
\eea

We can now give the perturbed version of the distortion tensor in terms of the metric potentials. Because this is to be integrated down the past lightcone, we write it in terms of $\d/\d\chi$ before other derivatives using the commutation relations. 
The Ricci contribution to the Hessian becomes [indices on the left are with respect to the full basis, on the right the conformal one]:
\bea
(1+z)^{-5}\mathcal{T}^{{}^\text{Ricci}}_{ABC}&=&
-\frac{1}{2}\delta_{A(B}\left\{\nabla_{C)}\nabla_D\nabla^D\Psi-
\HH\left(2\frac{\d}{\d\chi}+2\partial_\eta+3\HH\right)\nabla_{C)}\Phi
+\left[\frac{\d^2}{\d\chi^2}-\partial_\eta^2+\frac{2}{\chi}\frac{\d}{\d\chi}+\left(\frac{2}{\chi}-3\HH\right)\partial_\eta
\right]\nabla_{C)}\Psi      
\right\}\nonumber\\&&
-\frac{1}{2}\delta_{A(B}\nabla_{C)}\left[
\Psi''+\HH(2\Psi'+\Phi')+(2\HH'+\HH^2)\Phi
\right]\nonumber\\&&
-\frac{1}{4}\bigg\{
-\delta_{A(B}\nabla_{C)}\nabla_D\nabla^D
+\delta_{A(B}\left[\frac{\d^2}{\d\chi^2}+\left(-4\HH+\frac{3}{\chi}+\partial_\eta\right)\frac{\d}{\d\chi}
+\left(-4\HH+\frac{3}{\chi}\right)\partial_\eta\right]\nabla_{C)}
\nonumber\\&&
+\delta_{BC}\left[
-\frac{\d^2}{\d\chi^2}
+\left(4\HH-\frac{3}{\chi}-\partial_\eta\right)\frac{\d}{\d\chi}
+\left(4\HH-\frac{3}{\chi}\right)\partial_\eta
\right]\nabla_A
\bigg\}(\Psi-\Phi)\,.
\eea
We have presented this such that in the case of LCDM and GR, only the first line remains. 
The Weyl contribution becomes
\bea
(1+z)^{-5}\mathcal{T}^{{}^\text{Weyl}}_{ABC}&=&
\frac{1}{2}\bigg\{
-\nabla_{(B}\nabla_{C)}\nabla_A+\frac{1}{2}\delta_{A(B}\nabla_{C)}\nabla_D\nabla^D+\Gamma^D_{~~(BC)}\nabla_{A}\nabla_D+2\Gamma^D_{~~A(B}\nabla_{C)}\nabla_D
\nonumber\\&&
+\bigg[
-\frac{1}{2}\frac{\d^2}{\d\chi^2}+\left(\frac{9}{4}\HH-\frac{5}{4\chi}-\frac{1}{2}\partial_\eta\right)\frac{\d}{\d\chi}+\left(\frac{9}{4}\HH-\frac{5}{4\chi}\right)\partial_\eta+\frac{1}{\chi^2}
\bigg]\delta_{A(B}\nabla_{C)}
\nonumber\\&&
+\bigg[
+\frac{1}{2}\frac{\d^2}{\d\chi^2}-\left(\frac{7}{4}\HH-\frac{7}{4\chi}-\frac{1}{2}\partial_\eta\right)\frac{\d}{\d\chi}-\left(\frac{7}{4}\HH-\frac{7}{4\chi}\right)\partial_\eta+\frac{1}{\chi^2}
\bigg]\delta_{BC}\nabla_{A}
\nonumber\\&&
-\frac{3}{\chi}\left(\frac{\d}{\d\chi}+\frac{1}{\chi}+\partial_\eta\right)\delta_{(BC}\nabla_{A)}
-\left[\Gamma^D_{~~EA}\Gamma^E_{~~(BC)}+\Gamma^D_{~~E(C}\Gamma^E_{~~|B|A)}-\nabla_{(B}\Gamma^D_{~~|A|C)}\right]
\nabla_D
\nonumber\\&&
+
\frac{1}{\chi}\Gamma_{A(BC)}\left(\frac{\d}{\d\chi}+\frac{1}{\chi}\right)
\bigg\}(\Psi+\Phi)
\eea

As the final part of the solution we construct the integral projection operator. This is simple because $\mathcal{T}_{ABC}$ is already first order, so we just require it in the background where it reduces to the usual lensing kernel with small modifications. The  area distance in the background  is given in terms of the comoving distance   
\be
\bar d_A(\chi) = \frac{\chi}{1+z} = \frac{1}{(1+{z})}\int_{0}^{{z}} \frac{\d z'}{(1+z')\HH(z')}~~~\Rightarrow~~~\bar \theta = -\frac{2}{\chi}(1+z)^2(1-\mathcal{H}\chi)
\ee
from which we derive the reciprocal distance
\bea
\tilde d_A(\lambda)&=& \bar d_A(\lambda)\lim_{\epsilon\to0^+}
\left[\frac{1}{\bar d_A(\lambda_o-\epsilon)}+\int_{\lambda_o-\epsilon}^\lambda\frac{\d\lambda'}{{\bar d}_A(\lambda')^2}
\right]\\
&=& \frac{\chi}{1+z}\lim_{\epsilon\to0^+}
\left[\frac{1}{\epsilon}-\int_{\epsilon}^\chi\d\chi'\frac{1}{{\chi'}^2}
\right]=\frac{1}{1+z}\,.
\eea
 The projection operator simplifies to in FLRW,
\bea
\P&=&\I\int_{\lambda_o}^\lambda\d\lambda'\bar d_A(\lambda')^2\left[
\tilde d_A(\lambda)\bar d_A(\lambda')-
\bar d_A(\lambda)\tilde d_A(\lambda') 
\right]\\
&=&\frac{\I}{1+z}\int_0^\chi\d\chi'\frac{{\chi'}^2(\chi-\chi')}{(1+z(\chi'))^5}\,.
\eea
Finally we can give the full expression for the Hessian in the case of perturbed FLRW:
\be
\HH_{ABC}=\frac{1}{1+z}\int_0^\chi\d\chi'{{\chi'}^2(\chi-\chi')}\left[\frac{
\mathcal{T}^{{}^\text{Ricci}}_{ABC}(\chi')+\mathcal{T}^{{}^\text{Weyl}}_{ABC}(\chi')}{(1+z(\chi'))^5}\right]\,.
\ee
Note that the redshift terms cancel inside the integral with all integrated terms conformal comoving ones.

\subsection{Dominant contribution}

The dominant contribution to the distortion tensor arrises from terms with the highest number of screen-space derivatives, which are enhanced compared to radial derivatives on small scales. In this limit it is straightforward to write down the Hessian:
\bea
\mathcal{H}_{ABC}&=&\frac{1}{2(1+z)}\int_0^\chi\d\chi'{\chi'}^2(\chi'-\chi)\left[
\nabla_{(B}\nabla_{C)}\nabla_A(\Phi+\Psi)
\right]\\
&=& \frac{\chi^2}{2(1+z)}
\left[
\nabla_{(B}\nabla_{C)}\nabla_A\right]
\int_0^\chi\d\chi'\frac{\chi(\chi'-\chi)}{{\chi'}}(\Phi+\Psi)\\
&=& \frac{\chi^2}{2(1+z)}
\nabla_{(B}\nabla_{C)}\nabla_A\psi(\chi)
\eea
The invariant parts for the dominant part now become
\bea
\HA_A&=&\frac{3}{4}\frac{\chi^2}{2(1+z)}\nabla_A\nabla_B\nabla^B\psi\\
\HB_A&=&0\\
\widehat\HH_{ABC}&=&\frac{\chi^2}{2(1+z)}\nabla_{\<A}\nabla_{B}\nabla_{C\>}\psi\,.
\eea
Angled brackets in the last line denote the trace-free part. The spin 1 modes associated with rotation are not excited in this approximation, while the others are sourced directly by the density contrast. The spin 3 mode is sourced purely by the distortion of the electric Weyl curvature. 

\subsection{Modified Gravity}

In principle, measuring the Hessian allows us to reconstruct $\Phi$ and $\Psi$ separately. Equality of these two potentials is a clear test of GR. Crucially, the pure trace-free mode of the Hessian $\widehat\HH_{ABC}$ depends only on $\Phi+\Psi$, and is only induced by the Weyl distortion. However, the other two modes depend on $\Phi+\Phi$ but also on $\Psi$ and $\Phi$ in different ways. The Weyl contribution always gives $\Phi+\Psi$, but the Ricci part sources  the invariant parts of $\mathcal{T}^{{}^\text{Ricci}}_{ABC}$ as:
\bea
\frac{4}{3}(1+z)^{-5}\widehat\mathcal{T}^{{}^\text{Ricci}}_{A} &=&
-\frac{1}{4}\nabla_A\nabla_B\nabla^B(\Phi+\Psi)
\nonumber\\&&
+\left[
-\frac{11}{2}\frac{\d^2}{\d\chi^2}
+\left(9\HH-\frac{31}{4\chi}-\frac{1}{4}\partial_\eta\right)\frac{\d}{\d\chi}
+\left(\frac{19}{2}\HH-\frac{31}{4\chi}\right)\partial_\eta
\right]\nabla_A\Psi
\nonumber\\&&
+\left[
+\frac{9}{8}\frac{\d^2}{\d\chi^2}
+\left(-8\HH-\frac{21}{4\chi}+\frac{9}{4}\partial_\eta\right)\frac{\d}{\d\chi}
+\left(-\frac{17}{2}\HH+\frac{27}{4\chi}\right)\partial_\eta
+2\HH'+\frac{5}{2}\HH^2
\right]\nabla_A\Phi\,,
\eea
while the rotational invariant part becomes
\bea
3(1+z)^{-5}\overline{\mathcal{T}}^{{}^\text{Ricci}}_{A} &=&
+\left[
-\frac{3}{2}\frac{\d^2}{\d\chi^2}
+\left(-7\HH+\frac{4}{\chi}+\frac{3}{4}\partial_\eta\right)\frac{\d}{\d\chi}
+\left(-\frac{13}{2}\HH+\frac{17}{4\chi}\right)\partial_\eta
\right]\overline\nabla_A\Psi
\nonumber\\&&
+\left[
-\frac{23}{8}\frac{\d^2}{\d\chi^2}
+\left(+8\HH-\frac{69}{4\chi}-\frac{7}{4}\partial_\eta\right)\frac{\d}{\d\chi}
+\left(+\frac{15}{2}\HH-\frac{21}{4\chi}\right)\partial_\eta
+2\HH'+\frac{5}{2}\HH^2
\right]\overline\nabla_A\Phi\,,
\eea
Consequently, careful comparison of the 3 independent modes can give new tests of modified gravity.

\section{Conclusions}

We have derived for the first time the general solution to the Bazanski equation, which is the extension to the geodesic deviation equation which underlies weak lensing flexion. Consequently we have given the general equations for flexion in a perturbed FLRW model. An additional mode is found which depends on radial derivatives of the potentials along the line of sight. It may be used to test for modifications of GR, as it depends on the potentials differently from the other two flexion modes.

\acknowledgments 

I would like to thank David Bacon for discussions. This work is funded by the National Research Foundation (South Africa).


\begin{thebibliography}{99}

\bibitem{Seitz:1994xf} 
  S.~Seitz, P.~Schneider and J.~Ehlers,
  Class.\ Quant.\ Grav.\  {\bf 11}, 2345 (1994)
  [astro-ph/9403056].


\bibitem{Perlick:2004tq} 
  V.~Perlick,
  Living Rev.\ Rel.\  {\bf 7}, 9 (2004).


\bibitem[Bazanski(1977)]{1977AnIHP..27..145B} Bazanski, S.~L.\ 1977, 
Annales de L'Institut Henri Poincare Section Physique Theorique, 27, 145 


\bibitem[Bazanski(1977)]{1977AnIHP..27..115B} Bazanski, S.~L.\ 1977, 
Annales de L'Institut Henri Poincare Section Physique Theorique, 27, 115 






\bibitem{schutz}
Schutz, B. F. (1985). On Generalised Equations of Geodesic Deviation. In M. A. H. MacCallum (Ed.), Galaxies, axisymmetric systems and relativity: essays presentd to W. B. Bonnor on his 65th birthday (pp. 237-246). Cambridge: Cambridge University Press.

\bibitem{Vines:2014oba} 
  J.~Vines,
  arXiv:1407.6992 [gr-qc].



\bibitem{Goldberg:2004hh} 
  D.~M.~Goldberg and D.~J.~Bacon,
  Astrophys.\ J.\  {\bf 619}, 741 (2005)
  [astro-ph/0406376].

\bibitem{Bacon:2005qr} 
  D.~J.~Bacon, D.~M.~Goldberg, B.~T.~P.~Rowe and A.~N.~Taylor,
  Mon.\ Not.\ Roy.\ Astron.\ Soc.\  {\bf 365}, 414 (2006)
  [astro-ph/0504478].

\bibitem{Castro:2005bg} 
  P.~G.~Castro, A.~F.~Heavens and T.~D.~Kitching,
  Phys.\ Rev.\ D {\bf 72}, 023516 (2005)
  [astro-ph/0503479].
 
\bibitem{Okura:2006fi} 
  Y.~Okura, K.~Umetsu and T.~Futamase,
  Astrophys.\ J.\  {\bf 660}, 995 (2007)
  [astro-ph/0607288].

\bibitem{Goldberg:2006jp} 
  D.~M.~Goldberg and A.~Leonard,
  Astrophys.\ J.\  {\bf 660}, 1003 (2007)
  [astro-ph/0607602].

\bibitem{Massey:2006du} 
  R.~Massey, B.~Rowe, A.~Refregier, D.~J.~Bacon and J.~Berge,
  Mon.\ Not.\ Roy.\ Astron.\ Soc.\  {\bf 380}, 229 (2007)
  [astro-ph/0609795].
 
\bibitem{Schneider:2007ks} 
  P.~Schneider and X.~Er,
  Astron.\ Astrophys.\  {\bf 485}, 363 (2008)
  [arXiv:0709.1003 [astro-ph]].

\bibitem{Bernstein:2008ah} 
  G.~M.~Bernstein and R.~Nakajima,
  Astrophys.\ J.\  {\bf 693}, 1508 (2009)
  [arXiv:0807.1931 [astro-ph]].

\bibitem{Bacon:2008zj} 
  D.~J.~Bacon and B.~M.~Schaefer,
  Mon.\ Not.\ Roy.\ Astron.\ Soc.\  {\bf 396}, 2167 (2009)
  [Mon.\ Not.\ Roy.\ Astron.\ Soc.\  {\bf 396}, 2167 (2009)]
  [arXiv:0807.3663 [astro-ph]].

\bibitem{Lasky:2009ca} 
  P.~Lasky and C.~Fluke,
  Mon.\ Not.\ Roy.\ Astron.\ Soc.\  {\bf 396}, 2257 (2009)
  [arXiv:0904.1440 [astro-ph.CO]].

\bibitem{Bacon:2009aj} 
  D.~J.~Bacon, A.~Amara and J.~I.~Read,
  Mon.\ Not.\ Roy.\ Astron.\ Soc.\  {\bf 409}, 389 (2010)
  [arXiv:0909.5133 [astro-ph.CO]].

\bibitem{Leonard:2009fa} 
  A.~Leonard and L.~J.~King,
  Mon.\ Not.\ Roy.\ Astron.\ Soc.\  {\bf 405}, 1854 (2010)
  [arXiv:0910.0842 [astro-ph.CO]].

\bibitem{Hilbert:2010am} 
  S.~Hilbert, J.~R.~Gair and L.~J.~King,
  Mon.\ Not.\ Roy.\ Astron.\ Soc.\  {\bf 412}, 1023 (2011)
  [arXiv:1007.2468 [astro-ph.CO]].

\bibitem{Er:2010nt} 
  X.~Er, G.~Li and P.~Schneider,
  arXiv:1008.3088 [astro-ph.CO].

\bibitem{Pires:2010ar} 
  S.~Pires and A.~Amara,
  Astrophys.\ J.\  {\bf 723}, 1507 (2010)
  [arXiv:1009.0712 [astro-ph.CO]].

\bibitem{Munshi:2010cz} 
  D.~Munshi, T.~Kitching, A.~Heavens and P.~Coles,
  Mon.\ Not.\ Roy.\ Astron.\ Soc.\  {\bf 416}, 629 (2011)
  [arXiv:1012.3658 [astro-ph.CO]].

\bibitem{Fluke:2011ej} 
  C.~J.~Fluke and P.~D.~Lasky,
  Mon.\ Not.\ Roy.\ Astron.\ Soc.\  {\bf 416}, 1616 (2011)
  [arXiv:1101.4407 [astro-ph.CO]].

\bibitem{Schaefer:2011ui} 
  B.~M.~Schaefer, L.~Heisenberg, A.~F.~Kalovidouris and D.~J.~Bacon,
  Mon.\ Not.\ Roy.\ Astron.\ Soc.\  {\bf 420}, 455 (2012)
  [arXiv:1101.4769 [astro-ph.CO]].

\bibitem{Camera:2011me} 
  S.~Camera and A.~Diaferio,
  arXiv:1104.3955 [astro-ph.CO].

\bibitem{Rowe:2012ih} 
  B.~Rowe, D.~Bacon, R.~Massey, C.~Heymans, B.~Haeussler, A.~Taylor, J.~Rhodes and Y.~Mellier,
  Mon.\ Not.\ Roy.\ Astron.\ Soc.\  {\bf 435}, 822 (2013)
  [arXiv:1211.0966 [astro-ph.CO]].





\bibitem{Sachs:1961zz} 
  R.~K.~Sachs,
  Proc.\ Roy.\ Soc.\ Lond.\ A {\bf 264}, 309 (1961).

\bibitem[Jordan et al.(2013)]{2013GReGr..45.2691J} Jordan, P., Ehlers, J., 
\& Sachs, R.~K.\ 2013, General Relativity and Gravitation, 45, 2691 

\bibitem[Bertotti(1966)]{1966RSPSA.294..195B} Bertotti, B.\ 1966, Royal 
Society of London Proceedings Series A, 294, 195 

\bibitem{Clarkson:2011br}
  C.~Clarkson, G.~F.~R.~Ellis, A.~Faltenbacher, R.~Maartens, O.~Umeh and J.~P.~Uzan,
  Mon.\ Not.\ Roy.\ Astron.\ Soc.\  {\bf 426} (2012) 1121
  [arXiv:1109.2484 [astro-ph.CO]].

\bibitem{Ellis:1998ct} 
  G.~F.~R.~Ellis and H.~van Elst,
  NATO Sci.\ Ser.\ C {\bf 541}, 1 (1999)
  [gr-qc/9812046].

\bibitem{Maartens:1996ch} 
  R.~Maartens, G.~F.~R.~Ellis and S.~T.~C.~Siklos,
  Class.\ Quant.\ Grav.\  {\bf 14}, 1927 (1997)
  [gr-qc/9611003].


\bibitem{Clarkson:2002jz} 
  C.~A.~Clarkson and R.~K.~Barrett,
  Class.\ Quant.\ Grav.\  {\bf 20}, 3855 (2003)
  [gr-qc/0209051].
  
\bibitem{Clarkson:2007yp} 
  C.~Clarkson,
  Phys.\ Rev.\ D {\bf 76}, 104034 (2007)
  [arXiv:0708.1398 [gr-qc]].
    



\end{thebibliography}
\end{document}